\documentclass[aps,prl,superscriptaddress,showpacs,twocolumn]{revtex4-1}

\usepackage{graphicx}
\usepackage{amsmath}
\usepackage{xspace}
\usepackage{color}
\usepackage{bm,array}
\usepackage{dsfont}
\usepackage[colorlinks=true,linkcolor=blue,urlcolor=blue,citecolor=blue]{hyperref}

\newcommand{\ie}{i.e.\@\xspace}

\newcommand{\nag}{{\phantom{\dagger}}}

\newcommand{\eqw}[1]{(\ref{#1})}
\newcommand{\eq}[1]{Eq.\thinspace{}(\ref{#1})}
\newcommand{\eqq}[2]{Eqs.\thinspace{}(\ref{#1}) and (\ref{#2})}

\newcommand{\tab}[1]{Tab.\thinspace{}\ref{#1}}

\newcommand{\fig}[1]{Fig.\thinspace{}\ref{#1}}

\newcommand{\fc}[1]{({#1})}
\newcommand{\figc}[2]{Fig.\thinspace{}\ref{#1}\thinspace{}\fc{#2}}

\renewcommand{\d}{\ket{\downarrow}}
\newcommand{\up}{\ket{\uparrow}}

\def\bra#1{\mathinner{\langle{#1}|}}
\def\ket#1{\mathinner{|{#1}\rangle}}

\newcolumntype{R}{>{\raggedleft\arraybackslash}p{4em}}
\newcolumntype{S}{>{\raggedleft\arraybackslash}p{3em}}

\begin{document}

\title{Dissipative dynamics of a driven quantum spin coupled to a bath of ultracold fermions}

\author{Michael Knap}
\affiliation{Department of Physics, Harvard University, Cambridge MA 02138, USA}
\affiliation{ITAMP, Harvard-Smithsonian Center for Astrophysics, Cambridge, MA 02138, USA}
\author{Dmitry A. Abanin}
\affiliation{Department of Physics, Harvard University, Cambridge MA 02138, USA}
\affiliation{Perimeter Institute for Theoretical Physics, Waterloo, N2L2Y5 ON, Canada}
\author{Eugene Demler}
\affiliation{Department of Physics, Harvard University, Cambridge MA 02138, USA}

\date{\today}

\begin{abstract}
We explore the dynamics and the steady state of a driven quantum spin coupled to a bath of fermions, 
which can be realized with a strongly imbalanced mixture of ultracold atoms using currently 
available experimental tools. Radio-frequency driving can be used to induce tunneling between 
the spin states. The Rabi oscillations are modified due to the coupling of the quantum spin to the 
environment, which causes frequency renormalization and damping. The spin-bath coupling can be 
widely tuned by adjusting the scattering length through a Feshbach resonance. When the scattering 
potential creates a bound state, by tuning the driving frequency it is possible to populate either 
the ground state, in which the bound state is filled, or a metastable state in which the bound 
state is empty. In the latter case, we predict an emergent inversion of the steady-state 
magnetization. Our work shows that different regimes of dissipative dynamics can be 
explored with a quantum spin coupled to a bath of ultracold fermions.

\end{abstract}


\maketitle

Ultracold atomic systems provide a versatile laboratory to explore real-time many-body dynamics
due to their long coherence times and tunability~\cite{bloch_many-body_2008,ketterle_08} and, 
in particular, to study rich impurity physics~\cite{micheli_single_2004,recati_atomic_2005,bruderer_probing_2006,
zvonarev_ferrobosons_07,lamacraft_kondo_2008,lal_approaching_2010,orth_dynamics_2010,orth_universality_2010}.
In recent years, much progress has been achieved in realizing quantum impurities 
interacting with many-body environments. Examples include quantum degenerate gases 
consisting of a single atom type, where a few atoms are transferred 
to a different hyperfine state~\cite{schirotzek_observation_2009, nascimbene_collective_2009,palzer_quantum_2009,navon_equation_2010,fukuhara_quantum_2013},
ions immersed in quantum gases~\cite{zipkes_trapped_2010,schmid_dynamics_2010,ratschbacher_dynamics_2013},
and strongly imbalanced mixtures of multiple atomic species~\cite{will_coherent_2011,kohstall_metastability_2012,koschorreck_attractive_2012,catani_quantum_2012}.
Impurity atoms generally have several hyperfine states that interact 
differently with the host particles.  Coherent control of these 
states allows one to probe the influence of environmental coupling on impurity dynamics, e.g. in Fermi polarons, which 
are impurities dressed by particle-hole pairs~\cite{lobo_normal_2006,chevy_universal_2006,
chevy_ultra-cold_2010,schmidt_excitation_2011,massignan_repulsive_2011}. So far experimental studies focused mostly on the  spectral 
properties~\cite{schirotzek_observation_2009, kohstall_metastability_2012,koschorreck_attractive_2012} 
and mass renormalization~\cite{nascimbene_collective_2009,navon_equation_2010} of polarons. However, 
very recently Rabi oscillations of moving quantum spins, encoded in
two hyperfine states of the impurity atoms, have been
explored~\cite{kohstall_metastability_2012}, which gave further insights 
into polaron dynamics.

\begin{figure}
\begin{center}
 \includegraphics[width=0.49\textwidth]{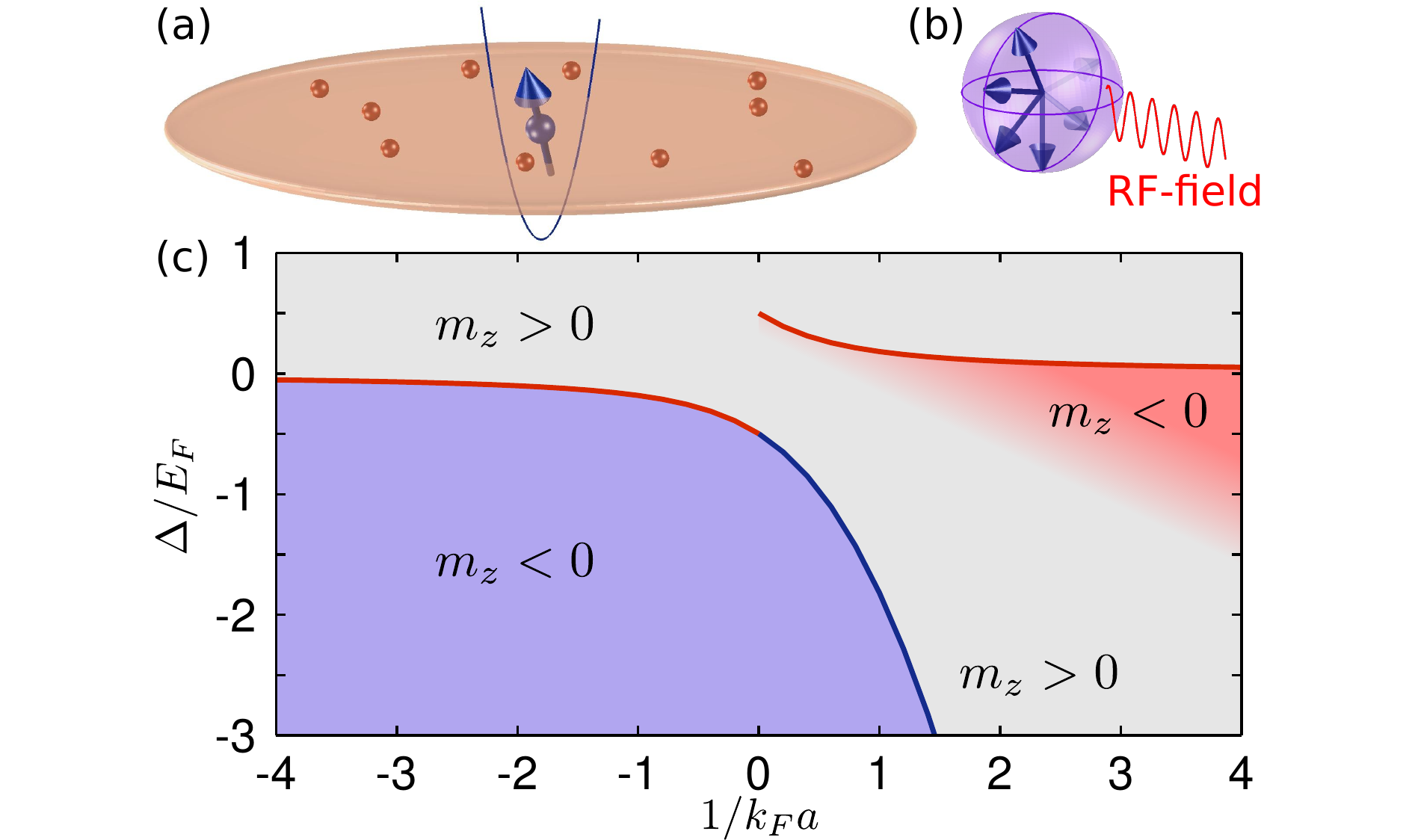}
\end{center}
\caption{\label{fig:co} (Color online) \fc{a} A driven quantum spin, which is dissipatively coupled to
a bath of fermions, can be experimentally realized with localized impurity atoms (sphere with arrow)
that are immersed in a Fermi gas (small spheres). \fc{b} Two hyperfine states of the 
impurities are driven by RF fields of strength $\Omega_0$ detuned from the bare 
transition by $\Delta$. \fc{c} The sign of the steady-state magnetization $m_z$ of the driven 
quantum spin is shown as a function of inverse scattering length $1/k_F a$ and
detuning $\Delta/E_F$. Along the solid lines $m_z$ vanishes and
non-trivial powerlaw frequency renormalizations and damping of the Rabi oscillations are found.
A transition from normal $m_z>0$ to inverted $m_z<0$ magnetization emerges below the 
metastable state, red shaded area. 
}
\end{figure}

Inspired by the tremendous experimental progress, here we analyze the dynamics of a 
quantum spin interacting with an ultracold fermionic bath (\figc{fig:co}{a)--(b}). 
We consider a situation where the spin performs radio-frequency (RF) driven Rabi oscillations, 
and study their frequency renormalization and damping due to spin-bath interactions. 
We find that this system realizes various regimes of dissipative dynamics. 
Moreover,  we predict an emergent inversion of the steady-state magnetization when
the scattering potential creates a bound state yet the driving is tuned to a metastable
state in which the bound state remains unoccupied.

We consider an experimentally relevant situation where the host fermions interact with 
the impurity via contact interactions, such that only $s$-wave scattering is important.
Without loss of generality, we assume that only one of the spin states, $\up$, 
interacts with host fermions with scattering length $a$, while the other, $\d$, does not. 
This scenario has been experimentally realized e.g. in \cite{kohstall_metastability_2012,koschorreck_attractive_2012}.
We mostly focus on the case when the quantum spin can be treated as immobile 
(this is true when the impurity is localized by a strong potential~\cite{loc} 
or is very heavy compared to host fermions). 
Two cases should be distinguished: (1) $a<0$, when the impurity potential does not create a 
bound state, and (2) $a>0$, when a bound state exists~\cite{combescot_1971,knap_time-dependent_2012}.
We show below through analytic arguments and numerical simulations 
that in both cases, in the low-energy limit, our problem maps onto 
the spin-boson model with an ohmic bath, characterized by the low-energy spectral 
density $J(\omega) = 2 \alpha \omega$~\cite{leggett_dynamics_1987,schon_quantum_1990,vojta_impurity_2006} where
$\alpha$ is the dimensionless coupling strength, that is widely tunable 
by changing the scattering length $a$. Further, the energy difference between the two 
states in the spin-boson model is controlled by the frequency of the driving field. 

In case (1), $a<0$, the dissipative coupling can be 
related to the scattering phase shift at the Fermi level, $\delta_F=-\tan^{-1}(k_F a)$, via 
\begin{equation}
\alpha_1=\frac{\delta_F^2}{2\pi^2}\,.
\label{eq:eta1}
\end{equation}
Even richer is case (2), $a>0$, where depending on the driving 
frequency, the physics of the driven quantum spin is
governed by effective spin-boson models with two different couplings. 
When the driving frequency is such that the bound state is populated 
during the Rabi oscillations, the coupling constant of the 
equivalent spin-boson model is 
\begin{equation}
\alpha_2=\frac{(\delta_F/\pi+1)^2}{2}.  
\label{eq:eta2}
\end{equation}
It is, however, also possible to tune the frequency to a {\it metastable} state
with unoccupied bound state. In that case, the coupling constant of the 
equivalent spin-boson model is given by $\alpha_1$, as in case (1).  
This allows one to explore a much broader range of coupling parameters, and, 
in particular, to approach the overdamped regime~\cite{leggett_dynamics_1987}.
Further, the dissipative phase transition of
the spin-boson model at $\alpha=1$~\cite{leggett_dynamics_1987} can
be explored with a multi-component Fermi bath.

\textbf{Model.---}Our system is described by an effective 
one-dimensional Hamiltonian 
\begin{equation}
 \hat{H} =  \hat H_0 + \ket{\uparrow}\bra{\uparrow} \otimes \hat V +  \Omega_0 \hat \sigma_x - \Delta \hat \sigma_z \;,
 \label{eq:h}
\end{equation}
where $\hat H_0 = \sum_k \epsilon_{k} ^\nag c_{k}^\dag c_{k}^\nag$ 
is the Hamiltonian of the host fermions and $\hat V = \frac{V}{L} \sum_{k,q} c_{k}^\dag c_{q}^\nag$ 
is the contact impurity scattering potential. The last two terms of \eq{eq:h} model the RF 
driving of the two impurity spin states $\d$, $\up$. The parameters $\Omega_0$ and $\Delta$ 
represent the tunneling amplitude between two spin states and the detuning, 
and can be independently controlled in experiment by changing the 
strength and frequency of the RF field. We will be interested in the situation where 
at $t<0$ the spin is in the $\d$ state, and fermions are in the ground state $\ket{\text{FS}}$. 
The driving is turned on at $t=0$. We will explore how the populations of the two 
spin states, $n_\sigma := \langle \hat n_\sigma(t) \rangle$, 
$\sigma \in \lbrace \downarrow, \uparrow \rbrace$, which
are readily accessible in experiments, evolve with time.

\begin{figure}
\begin{center}
 \includegraphics[width=0.49\textwidth]{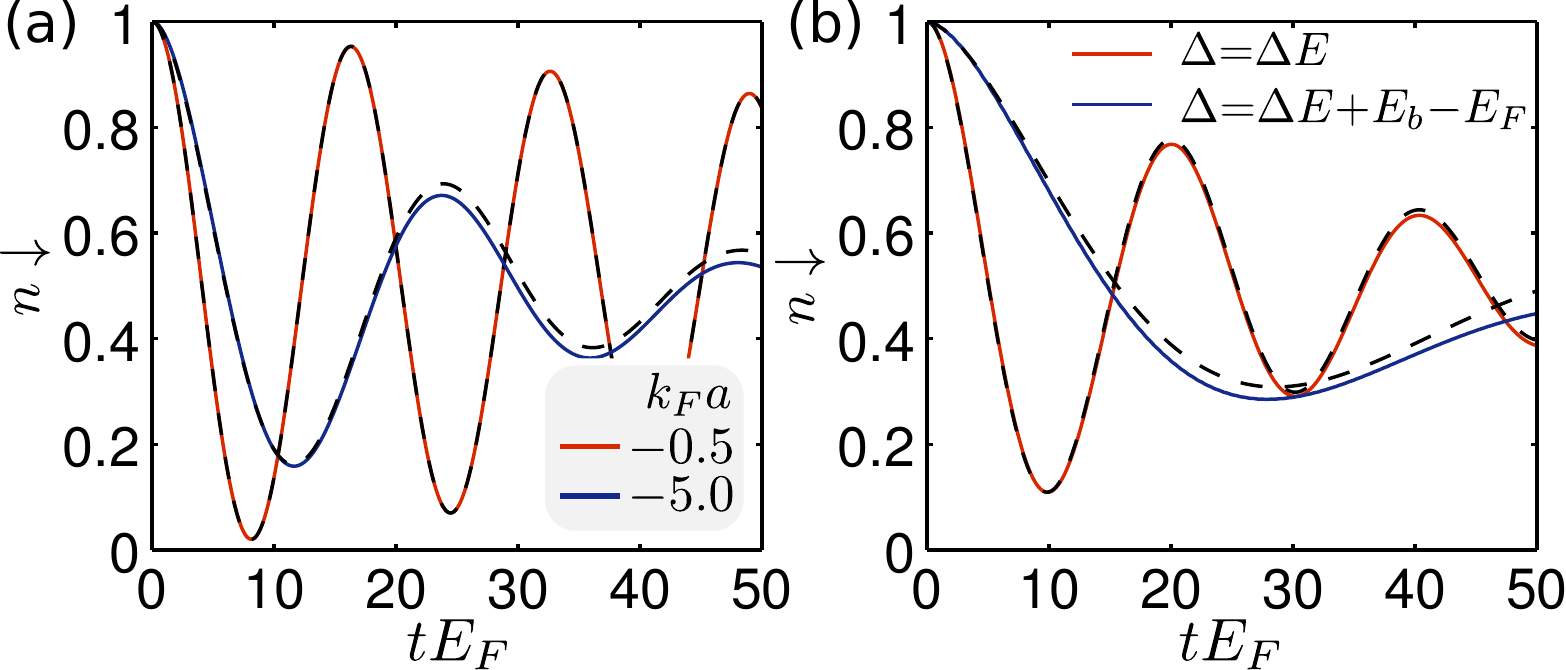}
\end{center}
\caption{\label{fig:rabiOsc} (Color online) Time dependent occupation $n_\downarrow$ of the 
state $\d$ at resonance $\epsilon=0$ for driving strength $\Omega_0=0.1 E_F$. The Rabi 
oscillations are strongly damped and their frequencies renormalized. \fc{a}
$n_\downarrow$ for negative scattering length $k_F a=\lbrace -0.5, -5.0\rbrace$. 
\fc{b} $n_\downarrow$ for the positive scattering length $k_F a=2$ and 
RF field tuned to $\Delta = \Delta E$ and to $\Delta = \Delta E + E_b -E_F$, 
respectively. In all cases the steady-state magnetization is zero. Solid lines are 
numerical results from the simulation of \eqw{eq:h} and dashed lines are obtained 
from perturbation theory. }
\end{figure}

\textbf{Relation to spin-boson model.---}In order to establish a  
low-energy description of our model for the case $a<0$ (no bound state),
we bosonize Hamiltonian \eqw{eq:h}~\cite{schotte_tomonagas_1969,guinea_bosonization_1985,giamarchi_quantum_2004,recati_atomic_2005,supp} and find
\begin{align}
 \hat H &= \sum_{q} v_F |q| b_q^\dag b_q^\nag  +  \sqrt{2\alpha_1}\pi v_F \sum_{q>0} \left( \frac{q}{2\pi L} \right)^{1/2} (b_q^\dag + b_q^\nag) \hat \sigma_z \nonumber\\
  &+   \Omega_0 \hat  \sigma_x + \epsilon \hat \sigma_z \;,
\label{eq:hbos}
\end{align}
which corresponds to the spin-boson model with an ohmic bath and dimensionless 
coupling $\alpha_1$. \eq{eq:eta1} relates $\alpha_1$ to the parameters of the microscopic model.

The energy of the $\up$ state is renormalized by the 
interactions with the Fermi sea \eqw{eq:h}~\cite{affleck_boundary_1997}: 
\begin{equation}
\label{eq:energy_shift}
\Delta E= -\int_0^{E_F} \frac{dE}{\pi} \delta(\sqrt{2mE}),
\end{equation}
where $m$ is the mass of host atoms and $E_F$ is the Fermi energy. 
Thus, the two families of states, one involving the $\d$ state and the 
other the $\up$ state, become effectively degenerate when the detuning 
compensates the energy renormalization, \ie, $\Delta=\Delta E$. 
Generally, the energy difference is $\epsilon=-\Delta+\Delta E$, which 
describes the effective bias of the quantum spin. 
The effective bias $\epsilon$, determines the steady-state magnetization
$m_z := \lim_{t\to\infty} \lbrace\langle \hat n_\uparrow(t)\rangle - \langle \hat n_\downarrow(t)\rangle\rbrace$~\cite{weiss_dynamics_1989,gorlich_low-temperature_1989}, whose sign we plot in \figc{fig:co}{c} as a function of the detuning $\Delta$ and 
the scattering length $a$. For $\epsilon=0$, \ie along the solid red line, 
 $m_z=0$.

For $a>0$, the situation is more involved, because 
of the presence of the bound state whose population dynamics influences the 
oscillations of the quantum spin. The oscillations occur between two families 
of states, one with the quantum spin in the $\d$ state and the other with 
the quantum spin in the $\up$ state. Our finding is that the latter can be 
either the ground state of the spin-up sector $|\Psi_{\uparrow}^{g} \rangle$ in 
which the bound state is occupied, or the metastable state $|\Psi_{\uparrow}^{m}\rangle$ 
in which the bound state is empty. 

In order to understand the two regimes and their properties, it is instructive 
to consider the correlation function $F(t)=\langle \Psi_{\downarrow}| e^{-i\hat Ht}|\Psi_{\downarrow}\rangle$. 
It should be noted that $F(t)$ determines spectral, 
rather than dynamical properties, yet, it will give us useful intuition. 
Following Yuval and Anderson~\cite{anderson_exact_1969,yuval_exact_1970}, 
$F(t)$ can be represented as a perturbative series in 
$\Omega_0$, where at order $n$ the spin flips $n$ 
times at times $t_1,\, t_2,\,\ldots\,t_n$. This reduces the problem 
to understanding the response of the Fermi gas to a potential introduced at $t_1,\,t_3,\,\ldots\,t_n$. 
In the absence of the bound state, such responses have a characteristic form of a Cauchy 
determinant~\cite{anderson_exact_1969}. The only parameter that enters 
those expressions is $\alpha_1$. To obtain $F(t)$, one then should sum 
over different spin flip times. 

When the impurity potential creates a bound state, the response of the Fermi gas for a 
given spin trajectory contains different contributions, coming from the intermediate 
states in which the bound state is either filled or empty. However, when detuning is 
such that $|\Psi_{\downarrow} \rangle$ is resonant with either the metastable 
$|\Psi_{\uparrow}^{m}\rangle$ or the ground state $|\Psi_{\uparrow}^{g}\rangle$, 
the contributions from intermediate states of one kind would dominate. Contributions 
of the other kind will oscillate rapidly (due to the large energy difference involved) 
and therefore, upon integration, their contribution will become negligible.  

For the case of only one spin flip, when the response function corresponds to the Anderson orthogonality catastrophe
~\cite{mahan_excitons_1967, 
anderson_infrared_1967, nozieres_singularities_1969} it is known that both contributions 
have a similar power-law form, but with different exponents. The first 
contribution (empty bound state) is characterized by an exponent $2\alpha_1$, 
while the second one (filled bound state) by
$2\alpha_2$. Generalizing this to the case 
of many spin flips, one can show, by extending the analysis of Combescot and 
Nozi\`eres~\cite{combescot_1971}, that the second contribution has the same form as the 
first one, but with exponent $2\alpha_2$. Thus, $F(t)$ is characterized by 
either $\alpha_1$ or $\alpha_2$ depending on the resonance condition. 

The above argument strongly suggests that, effectively, our model becomes 
equivalent to the spin-boson model with coupling $\alpha_2$ when 
$\Delta=\Delta E+E_b-E_F$, $E_b$ being the bound state energy, [blue line in \figc{fig:co}{c}] and with
$\alpha_1$ when $\Delta=\Delta E$, [red line in \figc{fig:co}{c}]. 
\begin{figure}
\begin{center}
 \includegraphics[width=0.49\textwidth]{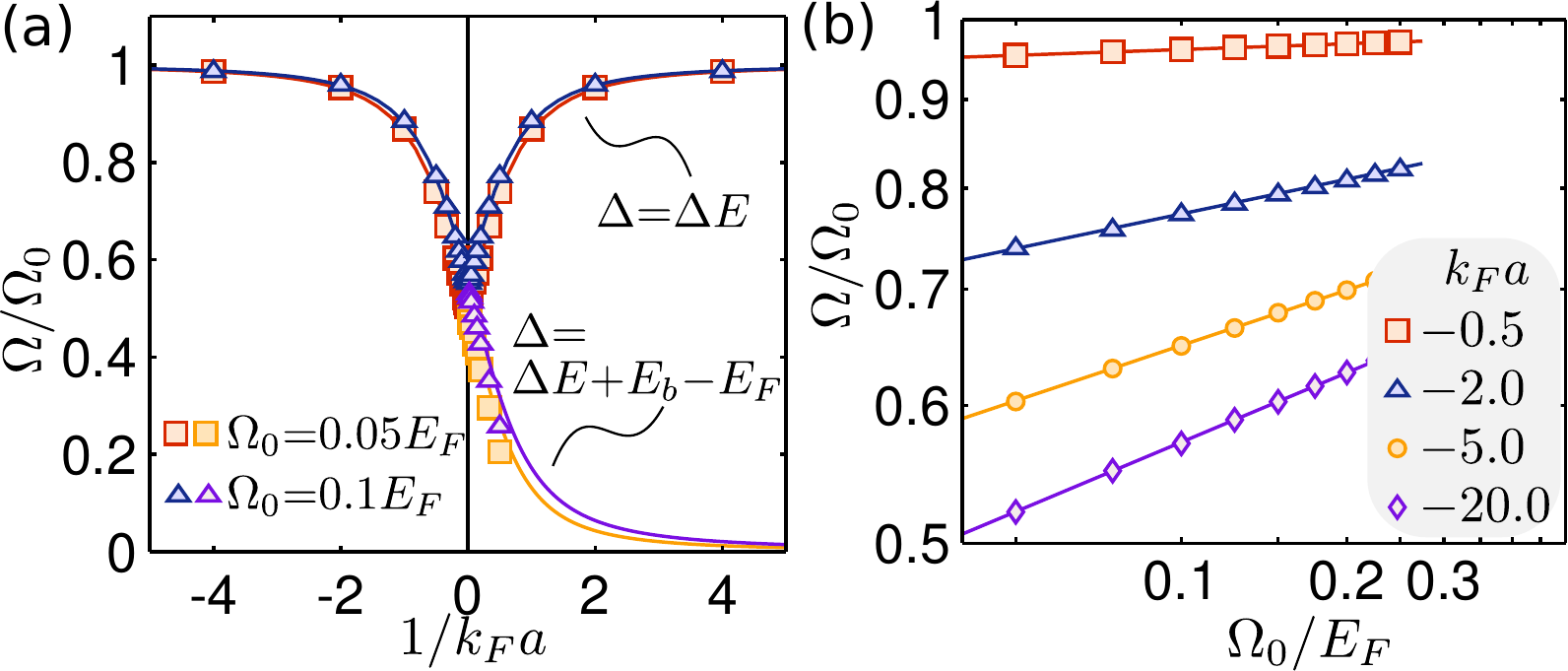}
\end{center}
\caption{\label{fig:rabiOmScaling} (Color online) Powerlaw renormalization of the
dressed Rabi frequency $\Omega_0$ at resonance $\epsilon=0$ due to coupling to 
the bath. \fc{a} Dressed Rabi frequency $\Omega/\Omega_0$ as a 
function of the inverse scattering length $1/k_F a$ for two different 
driving field strengths $\Omega_0=0.05 E_F$ and $\Omega_0=0.1 E_F$. 
\fc{b} Scaling of $\Omega/\Omega_0$ as a function of $\Omega_0$
for different values of the interaction strength $k_F a$.
The data is shown on a double logarithmic scale. The numerical results 
(symbols) are well described by analytic formula \eqw{eq:omDr} (solid lines).
}
\end{figure} 
To substantiate this expectation, we performed numerical 
simulations of the spin dynamics of Hamiltonian \eqw{eq:h} using  
matrix product states (MPS), where the initial ground state of the system is 
determined by density matrix renormalization group~\cite{white_density_1992,
schollwoeck_density-matrix_2005}. The time evolution with  
switched-on driving field  is calculated with time evolving block 
decimation~\cite{vidal_efficient_2003,vidal_efficient_2004}.
We choose in \eqw{eq:h} the dispersion of a one-dimensional 
lattice $\epsilon_k = -2J \cos k$ at half filling~\footnote{This
is similar to a continuum potential with effective range, which has also
been considered to describe experiment~\cite{kohstall_metastability_2012}.}, 
for which we find the relation $-k_F a = V/v_F$ by comparing the scattering phase shift of 
the lattice and the continuum.
We measure the occupation of the hyperfine states $\hat n_\downarrow $ and 
$\hat n_\uparrow $, respectively, from which we extract the renormalized Rabi 
frequency as well as damping by fitting to a damped, harmonic oscillator 
superimposed with a linear slope~\cite{supp}. 

\textbf{Driving at resonance $(\epsilon=0)$.---}We first consider zero effective detuning $\epsilon=0$
and thus follow the solid lines in \figc{fig:co}{c}, where $m_z=0$.
The numerically calculated time evolution of $n_\downarrow$ for driving strength 
$\Omega_0=0.1 E_F$, negative scattering length $a<0$, and $\Delta = \Delta E$ is 
shown in \figc{fig:rabiOsc}{a}, solid lines. With increasing
interaction strength $|k_F a|$, the Rabi frequency $\Omega$ is 
strongly reduced while the damping rate $\gamma$ is enhanced.
$n_\downarrow$ is shown in \fc{b} for positive scattering length $k_F a=2$ 
but different values of the detuning $\Delta=\Delta E$ and $\Delta = \Delta E + E_b - E_F$. 
When tuning to the bound state branch the dressed Rabi frequency
decreases significantly, illustrating that the coupling $\alpha$ increases 
due to the increase of scattering phase shift by $\pi$. 

To confirm the equivalence of the dynamics to that of the spin-boson model, 
we fit the numerical data by the analytical results obtained from noninteracting 
blip approximation (NIBA), which is a weak coupling expansion valid 
for $\alpha \ll 1/2$ and at short times~\cite{leggett_dynamics_1987,niba}.
Under NIBA the dynamics is divided into coherent and incoherent 
contributions. The coherent part consists of dressed Rabi oscillations 
of frequency $\Omega$ with a superimposed exponential damping $\gamma$ 
which are universally related through ${\Omega}/{\gamma} = -\tan {\pi}/(2-2\alpha)$. The dressed Rabi 
frequency $\Omega$ can be expressed as~\cite{leggett_dynamics_1987}
\begin{equation}
 \frac{\Omega}{\Omega_0} = F(\alpha) \left( \frac{\Omega_0}{\omega_c}\right)^\frac{\alpha}{1-\alpha} \;,
 \label{eq:omDr}
\end{equation}
with $F(\alpha):=\left[ \Gamma(1-2\alpha) \cos(\pi \alpha) \right]^{\frac{1}{2(1-\alpha)}} \sin \frac{\pi}{2(1-\alpha)}$ and 
$\omega_c$ is the high energy cutoff of the spin-boson theory. For 
$\Omega_0 < \omega_c$ this equation gives a \textit{reduction} of the dressed Rabi  
frequency $\Omega$ as compared to the driving strength $\Omega_0$. 
The dashed curves in \fig{fig:rabiOsc} are obtained from NIBA 
with the high energy cutoff as the only fitting parameter. For small interaction 
and at short to medium time scales NIBA describes the dynamics well. 

We extracted the renormalization of the Rabi frequency for several 
values of interaction strength and detuning for 
$\Omega_0=0.05 E_F$ and $\Omega_0=0.1 E_F$, \figc{fig:rabiOmScaling}{a}. The branch present 
for both positive and negative values of scattering length is obtained 
by setting $\Delta =\Delta E$, while the second branch at $a>0$ 
is obtained with $\Delta = \Delta E + E_b-E_F$. For small positive 
$k_Fa$ the driving cannot couple effectively to the bound state as its wavefunction
is of small spatial extend; hence for the ground state branch symbols 
are shown for $k_F a \gtrsim 1$. The powerlaw 
renormalization \eqw{eq:omDr} of the dressed Rabi frequency is demonstrated in \figc{fig:rabiOmScaling}{b}.
Thus we can conclude that the dynamics of our system is 
well-described by an effective spin-boson model for both positive and negative 
scattering length.

\begin{figure}
\begin{center}
 \includegraphics[width=0.49\textwidth]{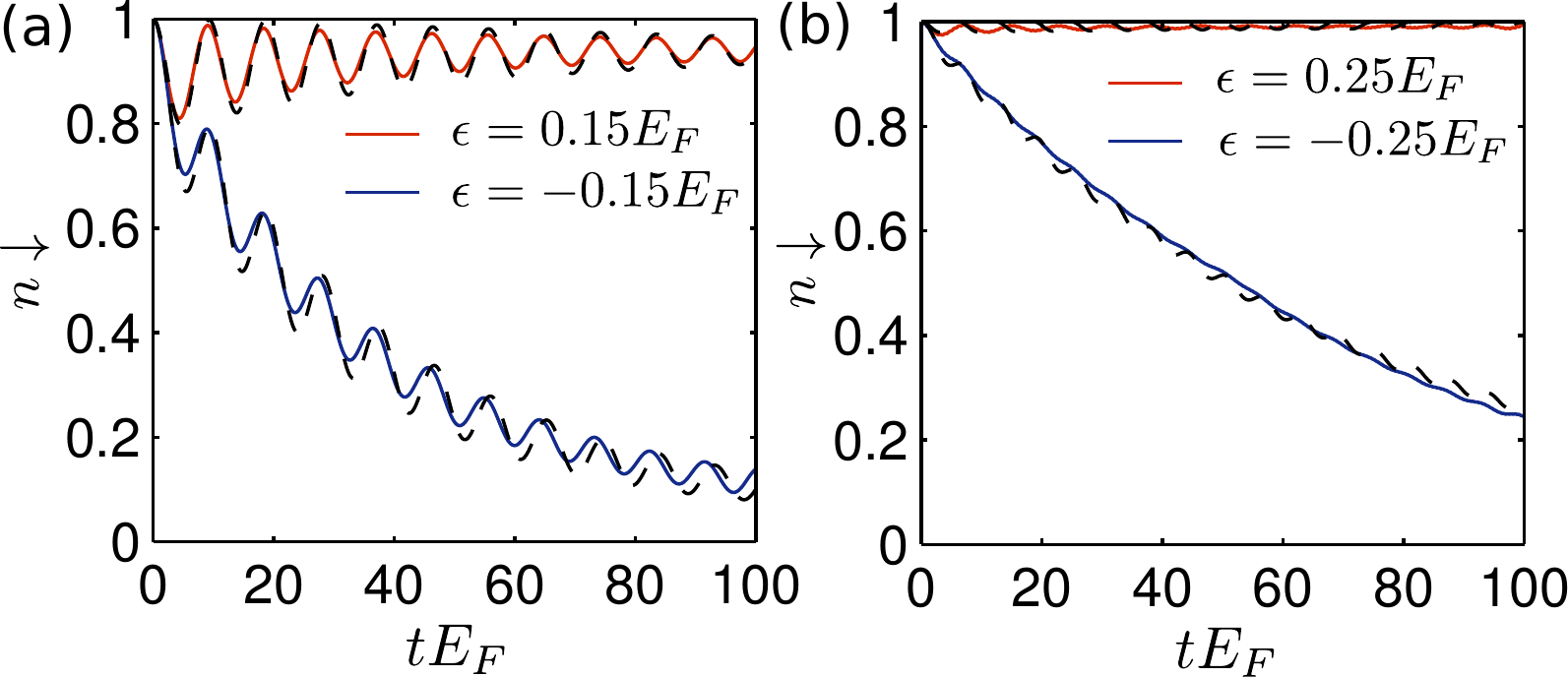}
\end{center}
\caption{\label{fig:rabiOscDet} (Color online) Time dependent occupation 
$n_\downarrow$ off resonance $\epsilon \neq 0$ for a driving 
strength $\Omega_0=0.1 E_F$ and scattering length \fc{a} $k_F a =-2$ and \fc{b} 
$k_F a =2$. For the latter the detuning is chosen with respect to the metastable branch. Solid lines 
are numerical and dashed lines perturbative results.}
\end{figure}

\textbf{Driving off resonance ($\epsilon \neq 0$).---}The case of off-resonant 
driving ($\epsilon \neq 0$) effectively corresponds to a biased spin-boson model.
Depending on the sign of $\epsilon$, the quantum spin has either positive 
or negative steady-state magnetization $m_z$, \figc{fig:co}{c}.

In \fig{fig:rabiOscDet} the numerically evaluated time-dependent occupation 
$n_\downarrow$, solid lines, is shown for $\Omega_0=0.1 E_F$, effective 
detuning $\epsilon=\pm 0.15 E_F$, and \fc{a} $k_F a=-2$ and \fc{b} $k_F a=2$. 
These numerical results are compared to a weak coupling expansion to first order in the 
blip-blip interaction~\cite{weiss_dynamics_1989,gorlich_low-temperature_1989}, dashed lines.
For negative scattering length \fc{a} and $\epsilon>0$ a quantum spin prepared in $\d$ decoheres 
only weakly; in agreement with $m_z<0$, \figc{fig:co}{c}. For $\epsilon<0$ ($m_z>0$), the occupation 
slowly flips with a rate that is indirect proportional to the detuning.
In \fc{b} the system is slightly detuned from the metastable branch. For 
$\epsilon<0$ the magnetization slowly reverts from negative to positive, indicating 
$m_z>0$, while for $\epsilon>0$, the $\d$ state remains 
highly occupied over long times, supporting the region of inverted magnetization below
the metastable branch shown in \figc{fig:co}{c}.

\textbf{Summary and discussion.---}We studied the dynamics of a driven quantum spin
coupled to a fermionic bath which can be realized with an imbalanced mixture of ultracold atoms.
Two hyperfine-states of the minority atoms serve as spin states and atoms themselves
are spatially localized by a strong optical lattice~\cite{loc}.
We used the mapping to the spin-boson model to study the problem analytically.
For the unbiased case ($\epsilon=0$), 
the spin-boson model exhibits a dissipative phase transition at coupling $\alpha=1$~\cite{leggett_dynamics_1987,schon_quantum_1990,vojta_impurity_2006}.  
With a single component bath, the coupling can take values %in the range 
\eqw{eq:eta1} $0<\alpha_1<1/8$ and \eqw{eq:eta2} $1/8 < \alpha_2 < 1/2$. 
For $a<0$ range \eqw{eq:eta1} can be explored, while
for $a>0$ both ranges are accessible.
To explore an even broader range of $\alpha$, one may consider an impurity immersed 
in a multi-component Fermi gas. Such gases can e.g. be realized with alkaline-earth 
atoms~\cite{fukuhara_degenerate_2007,taie_realization_2010,taie_su6_2012} 
that obey $\text{SU}(N)$ symmetry for which the coupling constant
is enhanced by $N$ compared to the single-channel case. 
Thus, for $N>2$ it should be possible to explore the dissipative phase transition.

Qualitatively, the existence of two resonances is reminiscent of 
experiments~\cite{kohstall_metastability_2012,koschorreck_attractive_2012}, which 
studied mobile impurities interacting with a three-dimensional Fermi gas. In this case, 
at $a>0$ stable and metastable polaron branches have been observed. 
Moreover, authors of Ref.~\cite{kohstall_metastability_2012} experimentally studied 
Rabi oscillations of the impurity spin at resonance. It should be noted, however, that for 
a mobile impurity, effectively, the Fermi gas provides a sub-ohmic, rather 
than ohmic bath. In this case the effect of the bath is generally described 
by damping, and no dissipative phase transition exists~\cite{leggett_dynamics_1987}. 
It is possible, however, that at short times, and for heavy 
impurities (as was the case in Ref.~\cite{kohstall_metastability_2012}), our results 
will still be applicable at least qualitatively.

The authors thank G.M. Bruun and E.G. Dalla Torre
for useful discussions and M. Ganahl for providing his MPS code~\cite{ganahl_observation_2012}.
The authors acknowledge support from Harvard-MIT CUA, the DARPA OLE program, 
AFOSR MURI on Ultracold Molecules, ARO-MURI on Atomtronics,
as well as the Austrian Science Fund (FWF) Project No. J 3361-N20. 
Numerical simulations have been performed on the Vienna Scientific Cluster.

%\bibliographystyle{apsrev4-1.bst}
%\bibliography{library,libraryadd}

%Merlin.mbs v4.21 2009-07-09.
%

\newpage
\appendix

\onecolumngrid
%\vspace{-1cm}

\begin{center}
 \large{\textbf{Supplemental material for:\\Dissipative dynamics of a driven quantum spin coupled to a bath of ultracold fermions}}
\end{center}

\section{Bosonization of the Hamiltonian}

Here, we provide details on the bosonization of Hamiltonian (3).
The low-energy description for the free fermion part is~\cite{giamarchi_quantum_2004}
\begin{equation}
  \hat H_0 = \sum_k \epsilon_k c_k^\dag c_k^\nag  \rightarrow \sum_k v_F k (c_{kR}^\dag c_{kR}^\nag- c_{kL}^\dag c_{kL}^\nag) =  \sum_k v_F k (c_{kR}^\dag c_{kR}^\nag+ c_{-kL}^\dag c_{-kL}^\nag )\;, 
\end{equation}
where we introduced left (L) and right (R) movers. We now consider new particles by performing the canonical transformation
\begin{subequations}
\label{eq:trans}
\begin{align}
 a_{kR} &= \frac{1}{\sqrt{2}} (c_{kR}+c_{-kL})\\
 a_{kL} &= \frac{1}{\sqrt{2}} (-c_{kR}+c_{-kL}) \;.
\end{align}
\end{subequations}
These new particles are fermions as they obey the respective anticommutation relations. 
Using the transformation \eq{eq:trans}, the free Hamiltonian reads
\begin{equation}
 \hat H_0  \rightarrow \sum_k v_F k (a_{kR}^\dag a_{kR}^\nag+ a_{kL}^\dag a_{kL}^\nag ) \;,
\end{equation}
which can be expressed as
\begin{equation}
  \hat H_0  \rightarrow \frac{v_F}{2\pi} \int dx (\nabla \theta)^2 + (\nabla \phi)^2 = \sum_q v_F |q| b_q^\dag b_q^\nag  \,. 
\end{equation}
using the standard bosonization prescription~\cite{giamarchi_quantum_2004}.

The transformation \eqw{eq:trans} combines positive and negative momenta of the original fermions. Therefore,
it would not be useful if generic quartic interactions were present in the Hamiltonian as those would be non-local in the
new operators~\cite{giamarchi_quantum_2004}. For our model, however, one finds that the interaction part between the fermions and the impurity 
couples only to the new right moving fermions
\begin{align}
  \hat V = \frac{V}{L} \sum_{k,q} c_k^\dag c_q^\nag \rightarrow  &\frac{V}{L} \sum_{k,q} (c_{kR}^\dag+c_{kL}^\dag) (c_{qR}^\nag + c_{qL}^\nag) =  
  \frac{2V}{L} \sum_{k,q} a_{kR}^\dag a_{kR}^\nag \nonumber \\&= -2V \frac{\nabla \phi_h(0)}{\pi} =
  -2 V \sum_{q>0} \left( \frac{q}{2\pi L} \right)^{1/2} (b_q^\dag + b_q^\nag)\,. 
  \label{eq:bosH}
\end{align}
In the second line we made use of the fact that right moving particles with full degrees of freedom
can be equivalently expressed as left and right moving particles with half as many degrees of freedom via
the relation $a_R(x<0)=a_L(-x)$~\cite{giamarchi_quantum_2004}. To indicate that only half of the degree's of freedom have to be considered
we added the subscript $h$ to the field $\phi(x)$. The full interaction part thus transforms as
\begin{equation}
 \ket{\uparrow}\bra{\uparrow} \hat V \rightarrow - V (\hat \sigma_z+\hat{\mathds{1}}) \frac{\nabla \phi_h(0)}{\pi}\,,
\end{equation}
where $\hat \sigma_z$ is shifted by the identity, since in the fermionic
model the interaction is proportional to $\ket{\uparrow}\bra{\uparrow}$. 
The local forward scattering $-V \frac{\nabla \phi_h(0)}{\pi}$ can be removed
by a transformation of the form
\begin{equation}
 \tilde \phi_h(x) = \phi_h(x) - \frac{V}{v_F} \Theta(x)\,,\qquad\tilde \phi_{\bar h}(x) = \phi_{\bar h}(x)\;,
 \label{eq:trans2}
\end{equation}
where $\Theta$ is the Heaviside step function. In \eq{eq:trans2} we transformed half of the degrees of 
freedom indicated by $h$ while the other half $\bar h$ remains invariant. With that one finds for the low 
energy properties of Hamiltonian (3)
\begin{align*}
 \hat H &\sim \frac{v_F}{2\pi} \int dx \left[(\nabla \theta)^2 + (\nabla \phi)^2 -\frac{2V}{v_F} (\hat \sigma_z+\hat{\mathds{1}}) \nabla \phi_h(x)\delta(x) \right] + \Omega_0 \hat \sigma_x - \Delta \hat \sigma_z \\
  &= \frac{v_F}{2\pi} \int dx \bigg[(\nabla \theta)^2 + (\nabla \tilde \phi)^2 +\frac{2V}{v_F} \nabla \tilde \phi_h(x) \delta(x)  + \frac{V^2}{v_F^2} \delta(x) \\
  &\qquad\qquad\qquad-\frac{2V}{v_F} (\hat \sigma_z+\hat{\mathds{1}}) \Big( \nabla \tilde \phi_h(x)\delta(x) + \frac{V}{v_F} \delta(x)\Big) \bigg] + \Omega_0 \hat \sigma_x - \Delta \hat \sigma_z  \\
  &=  \frac{v_F}{2\pi} \int dx \Big[(\nabla \theta)^2 + (\nabla \tilde \phi)^2 \Big] - V \hat \sigma_z \frac{\nabla \tilde \phi_h(0)}{\pi} + \Omega_0 \hat \sigma_x \underbrace{- (\frac{V^2}{\pi v_F} + \Delta)}_{= \epsilon} \hat \sigma_z + \text{const.}
\end{align*}

In standard boson notation the Hamiltonian reads
\begin{align}
 \hat H = \sum_q v_F |q| b_q^\dag b_q^\nag  +  V \sum_{q>0} \left( \frac{q}{2\pi L} \right)^{1/2} (b_q^\dag + b_q^\nag) \hat \sigma_z 
  +  \Omega \hat  \sigma_x + \epsilon \hat \sigma_z \;,
\label{eq:hbos}
\end{align}
which corresponds to Eq.~(4) in the main text when identifying 
\begin{equation}
\alpha = \frac{V^2}{2\pi^2v_F^2}\;.
\label{eq:alpha}
\end{equation}
We obtain the latter relation from comparison with the spectral-density of the spin-boson model~\cite{leggett_dynamics_1987}
\begin{equation}
 J(\omega)=\sum_q \lambda_q^2 \delta(\omega - v_F |q|) \;,
\end{equation}
where $\lambda_q$ describes the coupling to the bath. For an ohmic bath, the low-energy spectral function
is of the form 
\begin{equation}
 J(\omega) = 2\alpha \omega \;.
 \label{eq:ohmic}
\end{equation}
Comparing \eq{eq:hbos} with the spin-boson model~\cite{leggett_dynamics_1987}, we find
$\lambda_q= 2 V \left( \frac{q}{2\pi L} \right)^{1/2}$ for $q>0$ and $\lambda_q=0$ for $q<0$
and thus
\begin{align}
 J(\omega) = \frac{V^2}{\pi^2} \int_{0}^\infty dq\, q\, \delta(\omega - v_F |q|) = 2 \frac{V^2}{2 \pi^2 v_F^2} \omega\;.
 \label{eq:J}
\end{align}
From \eqq{eq:ohmic}{eq:J} we then find \eq{eq:alpha}.

\section{Numerical values}

In \tab{tab:values} we list numerical values for the ground state energy $E$, the spin-bath 
coupling $\alpha_1$, $\alpha_2$, and the energy renormalization $\Delta E$, $\Delta E + E_b-E_F$ 
as a function of the interaction strength $V$ ($k_Fa$) obtained from the lattice model,
which we simulate using matrix product states. The considered system consists of $L=200$ sites and $N=100$
particles.

\begin{table}[h]
 \caption{Numerical values for the ground state energy $E$, the spin-bath coupling 
$\alpha_1$, $\alpha_2$, and the energy renormalization $\Delta E$, $\Delta E + E_b-E_F$
as a function of the interaction strength $V$ ($k_Fa$). }
 \begin{tabular}{RSRSRSRSRSRSRS}
 \hline \hline
 
  $V\quad$	&&	$k_F a\quad$	&&	$E\quad$	&&	$\alpha_1\quad$	&&	$\Delta E\quad$	&&	$\alpha_2\quad$	& \multicolumn{3}{c}{$\Delta E + E_b - E_F$}	\\ \hline\hline
0.00	&&	0.00	&&	-127.462	&&	0.000	&&		&&		&&		&\\
0.10	&&	-0.05	&&	-127.413	&&	0.000	&&	0.0488	&&		&&		&\\
 0.20	&&	-0.10	&&	-127.367	&&	0.001	&&	0.0950	&&		&&		&\\
 0.30	&&	-0.15	&&	-127.323	&&	0.001	&&	0.1388	&&		&&		&\\
 0.40	&&	-0.20	&&	-127.282	&&	0.002	&&	0.1803	&&		&&		&\\
 0.45	&&	-0.23	&&	-127.262	&&	0.002	&&	0.2001	&&		&&		&\\
0.50	&&	-0.25	&&	-127.243	&&	0.003	&&	0.2193	&&		&&		&\\
0.60	&&	-0.30	&&	-127.206	&&	0.004	&&	0.2561	&&		&&		&\\
0.70	&&	-0.35	&&	-127.171	&&	0.006	&&	0.2907	&&		&&		&\\
0.80	&&	-0.40	&&	-127.139	&&	0.007	&&	0.3232	&&		&&		&\\
0.90	&&	-0.45	&&	-127.108	&&	0.009	&&	0.3537	&&		&&		&\\
1.00	&&	-0.50	&&	-127.080	&&	0.011	&&	0.3823	&&		&&		&\\
1.20	&&	-0.60	&&	-127.028	&&	0.015	&&	0.4343	&&		&&		&\\
1.40	&&	-0.70	&&	-126.982	&&	0.019	&&	0.4799	&&		&&		&\\
1.60	&&	-0.80	&&	-126.942	&&	0.023	&&	0.5201	&&		&&		&\\
1.80	&&	-0.90	&&	-126.906	&&	0.027	&&	0.5555	&&		&&		&\\
2.00	&&	-1.00	&&	-126.875	&&	0.031	&&	0.5868	&&		&&		&\\
4.00	&&	-2.00	&&	-126.696	&&	0.062	&&	0.7659	&&		&&		&\\
6.00	&&	-3.00	&&	-126.622	&&	0.079	&&	0.8402	&&		&&		&\\
10.00	&&	-5.00	&&	-126.558	&&	0.096	&&	0.9040	&&		&&		&\\
14.00	&&	-7.00	&&	-126.530	&&	0.103	&&	0.9322	&&		&&		&\\
20.00	&&	-10.00	&&	-126.508	&&	0.110	&&	0.9536	&&		&&		&\\
40.00	&&	-20.00	&&	-126.483	&&	0.117	&&	0.9787	&&		&&		&\\
60.00	&&	-30.00	&&	-126.475	&&	0.120	&&	0.9871	&&		&&		&\\
100.00	&&	-50.00	&&	-126.468	&&	0.122	&&	0.9938	&&		&&		&\\
-100.00	&&	50.00	&&	-226.468	&&	0.122	&&	-99.0062	&&	0.128	&&	1.0138	&\\
-60.00	&&	30.00	&&	-186.475	&&	0.120	&&	-59.0129	&&	0.130	&&	1.0204	&\\
-40.00	&&	20.00	&&	-166.483	&&	0.117	&&	-39.0213	&&	0.133	&&	1.0287	&\\
-20.00	&&	10.00	&&	-146.508	&&	0.110	&&	-19.0464	&&	0.141	&&	1.0533	&\\
-14.00	&&	7.00	&&	-140.530	&&	0.103	&&	-13.0678	&&	0.149	&&	1.0743	&\\
-10.00	&&	5.00	&&	-136.558	&&	0.096	&&	-9.0960	&&	0.158	&&	1.1020	&\\
-6.00	&&	3.00	&&	-132.622	&&	0.079	&&	-5.1598	&&	0.181	&&	1.1648	&\\
-4.00	&&	2.00	&&	-130.696	&&	0.062	&&	-3.2341	&&	0.210	&&	1.2380	&\\
-2.00	&&	1.00	&&	-128.875	&&	0.031	&&	-1.4132	&&	0.281	&&	1.4152	&\\
-1.80	&&	0.90	&&	-128.706	&&	0.027	&&	-1.2445	&&	0.294	&&	1.4462	&\\
-1.60	&&	0.80	&&	-128.542	&&	0.023	&&	-1.0799	&&	0.308	&&	1.4814	&\\
-1.40	&&	0.70	&&	-128.382	&&	0.019	&&	-0.9201	&&	0.324	&&	1.5213	&\\
-1.20	&&	0.60	&&	-128.228	&&	0.015	&&	-0.7657	&&	0.343	&&	1.5667	&\\
-1.00	&&	0.50	&&	-128.080	&&	0.011	&&	-0.6177	&&	0.363	&&	1.6184	&\\
-0.90	&&	0.45	&&	-128.008	&&	0.009	&&	-0.5463	&&	0.374	&&	1.6469	&\\
-0.80	&&	0.40	&&	-127.939	&&	0.007	&&	-0.4768	&&	0.386	&&	1.6773	&\\
-0.70	&&	0.35	&&	-127.871	&&	0.006	&&	-0.4093	&&	0.399	&&	1.7097	&\\
-0.60	&&	0.30	&&	-127.806	&&	0.004	&&	-0.3439	&&	0.412	&&	1.7442	&\\
-0.50	&&	0.25	&&	-127.743	&&	0.003	&&	-0.2807	&&	0.425	&&	1.7809	&\\
-0.45	&&	0.23	&&	-127.712	&&	0.002	&&	-0.2499	&&	0.432	&&	1.8001	&\\
-0.40	&&	0.20	&&	-127.682	&&	0.002	&&	-0.2197	&&	0.439	&&	1.8199	&\\
 -0.30	&&	0.15	&&	-127.623	&&	0.001	&&	-0.1612	&&	0.454	&&	1.8612	&\\
 -0.20	&&	0.10	&&	-127.567	&&	0.001	&&	-0.1050	&&	0.469	&&	1.9050	&\\
 -0.10	&&	0.05	&&	-127.513	&&	0.000	&&	-0.0512	&&	0.484	&&	1.9513	&\\
\hline\hline
 \end{tabular}

 \label{tab:values}
\end{table}

\end{document}